\documentclass[preprint]{elsarticle}

\usepackage{graphicx}
\usepackage{amssymb}
\usepackage{amsmath}
\usepackage{soul}

\usepackage{lineno}

\usepackage{setspace}
\usepackage{verbatim}
\usepackage{subfigure}
\usepackage{bbm}
\usepackage{textcomp}
\usepackage{multirow}
\usepackage[margin=2.5cm]{geometry}
\usepackage[skip=0.5\baselineskip]{caption}
\usepackage{array}
\newcolumntype{L}{>{\centering\arraybackslash}m{3cm}}
\newcolumntype{P}[1]{>{\centering\arraybackslash}p{#1}}
\newcolumntype{M}[1]{>{\centering\arraybackslash}m{#1}}

\usepackage[english]{babel}
\usepackage{graphicx}
\usepackage{chemfig,natbib}
\usepackage{framed}
\usepackage[normalem]{ulem}
\usepackage{amsmath}
\usepackage{amsthm}
\usepackage{amssymb}
\usepackage{amsfonts}
\usepackage{enumerate}
\usepackage{tabu}
\usepackage[version=4]{mhchem}
\usepackage[utf8]{inputenc}

\theoremstyle{definition}

\journal{Computers \& Chemical Engineering}

\begin{document}

\begin{frontmatter}


\title{A unified theory of emergent equilibrium phenomena in active and passive matter}


\author[add1]{Venkat Venkatasubramanian\textsuperscript{*}}
\ead{venkat@columbia.edu}

\author[add1]{Abhishek Sivaram}
\ead{as5397@columbia.edu}

\author[add1]{Laya Das}
\ead{ld2874@columbia.edu}

\address[add1]{Complex Resilient Intelligent Systems Laboratory, Department of Chemical Engineering, Columbia University, New York, NY 10027}

\begin{abstract}

Recent attempts towards a theory of active matter utilize concepts and methods from hydrodynamics, kinetic theory, and non-equilibrium statistical physics. However, such approaches typically do not seem to recognize the critical feature of some kinds of active matter (particularly the biological ones), namely, the role of \textit{purpose}, and the naturally attendant concept of the \textit{pursuit of maximum utility}, which we believe is the crucial difference between active and passive matter. Here we introduce a novel game-theoretic framework, \textit{statistical teleodynamics}, that accounts for this feature explicitly and show how it can be integrated with conventional statistical mechanics to develop a unified theory of \textit{arbitrage equilibrium} in  active and passive matter. We propose a spectrum of self-actualizing capabilities, going from none to completely strategic decision-making, and envision the various examples of active matter systems occupying someplace in this spectrum. We show how statistical teleodynamics reduces to familiar results in statistical mechanics in the limit of zero self-actualization. At the other extreme, in an economic setting, it provides novel insights about the emergence of income distributions and their fairness in an ideal free-market society. As examples of agents in between these limits, we show how our theory predicts the behavior of active Brownian particles, the emergence of ant craters, and phase equilibria in social segregation dynamics. We suggest that our theory offers a novel systems theoretic perspective of emergent phenomena that could serve as the starting point for a more comprehensive theory of design, control, and optimization through self-organization. \\  
\end{abstract} 


\end{frontmatter}


\section*{Introduction}
Active matter describes systems composed of large numbers of self-actualizing entities or agents, which consume and dissipate energy resulting in interesting out-of-equilibrium system behavior \cite{marchetti2013hydrodynamics, toner2005hydrodynamics, narayan2007long, ramaswamy2010mechanics}. Biological examples of such systems include self-organizing bio-polymers, bacteria, schools of fish, and flocks of birds. Nonliving active matter examples include self-propelled particles, layers of vibrated granular rods, etc. A central conceptual puzzle in our evolving understanding of active matter is why and when does a collection of active entities that looks like a dynamic, out-of-equilibrium system at the microscopic scale, behave macroscopically like a simple equilibrium system of passive matter
\cite{berkowitz2020active, o2020lamellar, cates2015motility, cates2013active, gonnella2015motility}.\\

Recent advances towards a theory of such systems make use of concepts and methods drawn from hydrodynamics, kinetic theory, and non-equilibrium statistical physics, \cite{marchetti2013hydrodynamics, toner2005hydrodynamics, narayan2007long, ramaswamy2010mechanics, o2020lamellar}. However, such approaches do not typically  recognize \textit{explicitly} the critical feature of some active agents, particularly the biological ones, namely, the role of \textit{purpose} or \textit{goal} and the incessant \textit{pursuit of maximum utility} through self-actualization,  which we believe is the main difference between active and passive matter. By \textit{passive}, we mean that the entities are \textit{not goal-driven}. Gas molecules in a container, for example, do not pursue goals or make decisions in the paths they follow. They lack \textit{agency} and are mere prisoners of Newton's laws and conservation equations.\\

On the other hand, we consider the term active matter broadly and envision a spectrum of self-actualizing capabilities (i.e., agency), going from none (say, 0\%), as in the case of randomly moving about gas molecules, to completely strategic (say, 100\%) decision-making agents, somewhat like the rational \textit{Homo Economicus} of neoclassical economics. As a unifying conceptual framework, we propose that one might consider the various examples of active matter systems, such as a collection of self-driven particles, bacteria, fish, birds, and even humans, occupying some spot in this spectrum. While the entities at lower-levels of this spectrum behave \textit{without purpose} and therefore make \textit{random}, not strategic, choices, the role of purpose becomes more central as we move up the scale. The entities acquire more and more agency and begin to make \textit{strategic}, not random, decisions to maximize their utility or satisfaction. Therefore, any comprehensive theory of active matter has to recognize \textit{explicitly} this crucial and defining characteristic of active agents. \\

Here we describe such a theory, which we call \textit{statistical teleodynamics} \cite{venkatasubramanian2017statistical, venkat2017}. The name comes from the Greek word \textit{telos}, which means goal. Just as the dynamical behavior of gas molecules is driven by thermal agitation (hence, \textit{thermo}dynamics), the dynamics of purposeful agents is driven by the pursuit of their goals and, hence, \textit{teleo}dynamics. Statistical teleodynamics may be considered as the natural generalization of statistical thermodynamics for purpose-driven agents in active matter. It is a synthesis of the central concepts and techniques of potential games theory with those of statistical mechanics towards a unified theory of emergent equilibrium phenomena in active and passive matter. \\

Such a bold idea of a unified theory that encompasses both purpose-free and purposeful agents may seem preposterous. It is perfectly reasonable to be skeptical and ask how can the same theory correctly predict and explain the emergent behavior of purpose-free random molecules as well as that of purpose-driven people, which are at the opposite ends of the capabilities spectrum. However, we show that such a surprising universal theory is possible for certain types of emergent behavior. We show that this seemingly absurd idea has already yielded a valuable result by answering an important 200-year-old open-question in political economy.\\  

The rest of the paper is organized as follows. After a brief introduction to potential game theory, we show how the behavior of purpose-free matter can be seen as a limiting case of statistical teleodynamics through the example of the \textit{Thermodynamic Game} for molecules.  This may be considered as an example of the 0\%-limit. We then present the other extreme, the 100\%-limit, with rational humans as agents competing for jobs in an ideal free-market (the \textit{Income Game}) and show how our theory correctly predicts the emergence of income distributions in an ideal free-market society. \\

Having demonstrated the two extreme cases, we then present three case studies of very different kinds of agents in between, the emergent behavior of active Brownian particles (the \textit{Chemotaxis Game}), the collective behavior of ants building craters (the \textit{Ant Game}), and the segregation dynamics and phase equilibria of social agents (the \textit{Schelling Game}). Our work on the \textit{Thermodynamic Game} and the \textit{Income Game} has been published before ~\cite{venkat2015, venkat2017}, but we include them here to "bookend" the overall framework of the self-actualizing capabilities spectrum wherein the new results reported here (namely, active Brownian motion, ant craters formation, and social segregation dynamics) fit in. We conclude with a discussion of the main results and their implications. 


\section*{Population games}

The theory of population games addresses the following problem. Given a large collection of strategically interacting rational agents, where each agent is trying to decide and execute the best possible course of actions that maximizes the agent's~{\em payoff} or {\em utility} (we will use payoff and utility interchangeably in this paper; the term payoff is generally preferred in game-theoretic settings) in light of similar strategies executed by the other agents, we are interested in predicting what strategies would be executed and what outcomes are likely~\cite{easley2010networks, sandholm2010population}. In particular, one would like to know whether such a game would lead to an equilibrium situation.\\

Interestingly, it is the same question one has when dealing with a large number of entities that are {\em not} interacting strategically, but \textit{randomly}, such as molecules of a gas enclosed in a container. We want to know what the equilibrium state is for the gas and what criterion determines it. This is answered using statistical mechanics, of course. However, since both mathematical frameworks, game theory, and statistical mechanics, define and determine an equilibrium state, it is natural to ask whether there is any connection between the two approaches. As it turns out, there is a deep, beautiful connection with surprising and useful insights. \\

While it is obvious that rational individuals and many biological agents actively pursue goals and strategies, one might object that this is not the case for some active matter entities such as self-propelled particles.  We agree, but the latter behave in a manner - namely, the self-driving feature - that seems like goal-seeking behavior and so could be treated as such in our framework as we show below. \\

In our search for a unified theory, our goal is to identify the fundamental principles and develop simple models that offer an appropriate coarse-grained description of the system and make predictions not restricted by system-specific details and nuances. We have tried, deliberately, to keep the models as simple as possible without losing key insights and their relevance to empirical phenomena.\\

\subsection*{Population games and potential function}

Population game theory provides an analytical framework for studying the strategic interactions of a large population of agents with the following properties, as described by Sandholm~\cite{sandholm2010population}: (i) The number of agents is large; (ii) Individual agents play a small role - any one agent's behavior has only a small effect on other agents' payoff; (iii) Agents interact anonymously -- each agent's payoff only depend on opponents' behavior through the distribution of their choices; (iv) The number of roles is finite -- each agent is a member of one of a finite number of populations, and (v) Payoffs are continuous -- this property makes sure that very small changes in aggregate behavior do not lead to large changes in payoffs. We note that these properties more than adequately account for the features of our applications. \\
    
In games like Prisoner's Dilemma, with a small number of players, the typical game-theoretic analysis proceeds by systematically evaluating all the options every player has, determining their payoffs, writing down the payoff matrix, identifying the best responses of all players, identifying the dominant strategies (if present) for everyone, and then finally reasoning whether there exists a Nash Equilibrium (or multiple equilibria) or not. However, for population games, given a large number of players, this procedure is not feasible for several reasons -- for instance, a player may not know of all the strategies others are executing (or could execute), and their payoffs, for her to determine her best response.\\
    
Fortunately, as it turns out, for some population games, one can identify a single scalar-valued global function, called a {\em potential} ($\phi(\mathbf{x})$), that captures the necessary information about the payoffs {(where $\mathbf{x}$ is the state vector of the system)}. The {\em gradient} of the potential is the payoff or utility. Such games are called {\em potential games}~\cite{rosenthal1973class,sandholm2010population, easley2010networks, monderer1996potential}. A potential game reaches equilibrium, called \textit{Nash equilibrium}, when the potential $\phi(\mathbf{x})$ is maximized. Furthermore, this Nash equilibrium is unique if  $\phi(\mathbf{x})$ is strictly concave \cite{sandholm2010population}. \\

As noted, in potential games, a player's payoff or utility {for state $i$}, $h_i$, is the gradient of potential $\phi(\mathbf{x})$, i.e.,
\begin{equation}
{h}_i(\mathbf{x})\equiv {\partial \phi(\mathbf{x})}/{\partial x_i}
\end{equation}
where $x_i=N_i/N$ and $\mathbf{x}$ is the population vector. {$N_i$ is the number of agents in state $i$ and $N$ are the total number of agents}. Therefore, by integration (we replace partial derivative with total derivative because ${h}_i(\mathbf{x})$ can be reduced to ${h}_i(x_i)$), we have 
\begin{eqnarray}
\phi(\mathbf{x})&=&\sum_{i=1}^n\int {h}_i(\mathbf{x}){d}x_i \label{eq:potential}
\end{eqnarray}

where $n$ is the total number of states. \\

To determine the maximum potential, we use the method of Lagrange multipliers with $L$ as the Lagrangian and $\lambda$ as the Lagrange multiplier for the constraint $\sum_{i=1}^nx_i=1$:

\begin{equation}
L=\phi+\lambda(1-\sum_{i=1}^nx_i)\label{lagrangian}
\end{equation}\\
    
At equilibrium, all agents enjoy the same utility, $h_i = h^*$. In fact, the equality of utilities in all states is the \textit{fundamental criterion of game-theoretic equilibrium} for active matter. It is an \textit{arbitrage equilibrium} \cite{kanbur2020occupational} where the agents don't have any incentive to switch states anymore as all states provide the same utility $h^*$. In other words, equilibrium is reached when the opportunity for arbitrage, i.e., the ability to increase one’s payoff or utility by simply switching to another option or state at no cost, disappears. Thus, the maximization of $\phi$ and $h_i = h^*$ are exactly equivalent criteria, and both specify the same outcome, namely, an arbitrage (i. e., Nash) equilibrium. The former stipulates it from the \textit{top-down, system, perspective} whereas the latter the \textit{bottom-up,  agent,} perspective. \\

\section*{Thermodynamics of gas molecules as a game}	

With this quick introduction to potential games, we now show how the thermodynamic equilibrium of gas molecules in a container could be viewed from the perspective of population games. We call this the \textit{Thermodynamic Game}. This result has been published before \cite{venkat2015, venkat2017} and we include it here for the convenience of the reader. Furthermore, it is a natural place to start, the 0\%-"bookend" of the capabilities spectrum. \\

Now, as noted, gas molecules are purpose-free and hence don't chase after utility. However, in this game-theoretic formulation, we show that our agents, when they pursue a particular form of "utility" as given in Eq~\eqref{eq:thermo_payoff}, they behave like gas molecules. So, approaching this \textit{Thermodynamic Game} from the potential game perspective, we introduce the following "utility," $h_i$, for molecules in state $i$:

\begin{equation}
{h}_i(E_i,N_i)=-\beta E_i - \ln N_i\label{eq:thermo_payoff}
\end{equation}

where $E_i$ is the energy of a molecule in state $i$, $N_i$ is the number of molecules in state $i$, $\beta = 1/k_BT$, $k_B$ is the Boltzmann constant and $T$ is temperature. The first term models their tendency to prefer lower energy states {due to an increased utility in the lower energy state}. The $-\ln N_i$ term models the \textit{disutility of competition}. This term models the "restless" nature of molecules and their propensity to spread out. This is so because the $-\ln N_i$ term incentivizes the agent to leave its current location of higher $N_i$ to a location of lower $N_i$ all the time. \\ 

By integrating this utility{, and using the Stirling approximation for large number of particles $N_i$}, we can obtain the potential of the \textit{Thermodynamic Game}:

\begin{equation}
\phi(\mathbf{x})=-\frac{\beta}{N} E +\frac{1}{N} \ln\frac{N!}{\prod_{i=1}^n(Nx_i)!}\label{eq:thermo_potential}
\end{equation}

where $E=N\sum_{i=1}^nx_iE_i$ is the total energy that is conserved, $N$ is the total number of molecules, $n$ is the total number of states, and $x_i = N_i/N$.\\

This game reaches Nash equilibrium when $\phi(\mathbf{x})$ is maximized~\cite{sandholm2010population}. Furthermore, this equilibrium is unique as $\phi(\mathbf{x})$ is strictly concave:
\begin{eqnarray}
{\partial^2 \phi(\mathbf{x})}/{\partial x_i^2}=-{1}/{x_i}<0
\end{eqnarray}

To determine the actual equilibrium distribution, we use the method of Lagrange multipliers with $L$ as the Lagrangian and $\lambda$ as the Lagrange multiplier for the constraint $\sum_{i=1}^nx_i=1$:

\begin{equation}
L=\phi+\lambda(1-\sum_{i=1}^nx_i)\label{eq:lagrangian}
\end{equation}

Solving $\partial L/\partial x_i=0$ and substituting the results back in $\sum_{i=1}^nx_i=1$, we obtain the well-known {\em Boltzmann exponential distribution} of energy at equilibrium: 

\begin{equation}
x_i^*=\frac{\exp(-\beta E_i)}{\sum_{j=1}^n\exp(-\beta E_j)}\label{eq:boltzmann}
\end{equation}

Thus, this game's Nash equilibrium is the same as the statistical thermodynamic equilibrium, as expected. The critical insight here is, as Venkatasubramanian et al. ~\cite{venkat2015} first showed, that the second term in Eq~\eqref{eq:thermo_potential}, 

\begin{equation*}
\ln\frac{N!}{\prod_{i=1}^n(Nx_i)!}
\label{eq:entropy}
\end{equation*}

is the same as entropy (except for the Boltzmann constant). Thus, maximizing potential in population game theory is equivalent to maximizing entropy in statistical mechanics subject to constraints (which is the first term in Eq~\eqref{eq:thermo_potential}, the constraint on total energy $E$)~ \cite{jaynes1957information, jaynes1957information2, kapur1989maximum}. This is a deep and beautiful connection between statistical mechanics and potential game theory that we didn't know before. This fundamental connection allows us to generalize statistical thermodynamics to statistical teleodynamics and lays the foundation towards a universal theory of emergent equilibrium behavior of both passive and active agents.\\ 

Readers will recognize that from Eq~\eqref{eq:thermo_potential}, we have:

\begin{equation}
\phi=-\frac{1}{Nk_BT} (E-TS)=-\frac{\beta}{N} A
\end{equation}

where $A = E - TS$ is the Helmholtz free energy. Indeed, in statistical thermodynamics, $A$ is called a {\em thermodynamic potential}. Again, we see the correspondence between game-theoretic potential and thermodynamic potential.  In this regard, we could also see the correspondence between utility $h_i$ and {\em chemical potential}, the partial molar free energy, with an important difference. Active agents try to increase their utilities, whereas passive agents try to decrease their chemical potential.  In statistical teleodynamics, equilibrium is reached when all agents have the same utility, whereas, in statistical thermodynamics, it is reached when all agents have the same chemical potential. In this \textit{Thermodynamic Game} example, both lead to the same equilibrium, of course, as they are the same. \\

The statistical teleodynamics perspective reveals a deep but underappreciated insight in statistical mechanics, which is the recognition that when $N_i$ (or $x_i$) follows the exponential distribution in Eq~\eqref{eq:boltzmann}, $h_i = h^*$ for all the molecules. As noted, this is the defining criterion for game-theoretic or arbitrage equilibrium. Thus, the exponential distribution is a curve of \textit{constant effective utility} (or equivalently constant chemical potential), for all values of $E_i$. That is, it is an \textit{ \textit{isoutility}} curve or distribution for $h_i$ defined by Eq~\eqref{eq:thermo_payoff}. This turns out to be a particularly valuable insight in the context of fair income distribution as we discuss next. \\

Our approach reveals another valuable insight that is also not readily seen in statistical mechanics. We realize that we don't have to maximize the potential $\phi(x)$ necessarily to derive the equilibrium distribution. We can adopt the \textit{agent perspective} and recognize that equilibrium is reached when all agents enjoy the same utility, $h_i = h^*$. Therefore, we have 

\begin{equation}
{h}_i= {h^*} = -\beta E_i - \ln N_i^*, ~~i \in \{1,\dots,n\}
\label{eq:equil-h}
\end{equation}

where $N_i^*$ is given by the equilibrium distribution. From this, it is easy to derive the Boltzmann energy distribution by rearranging and solving for $N_i^*$ as

\begin{equation*}
x_i^*=\frac{N_i^*}{N} = (1/N)\exp{(-h^*)}\exp(-\beta E_i)
\end{equation*}

As before, by applying the constraint $\sum_{i=1}^nx_i=1$, we obtain the familiar form

\begin{equation}
x_i^*=\frac{\exp(-\beta E_i)}{\sum_{j=1}^n\exp(-\beta E_j)}\label{eq:boltzmann2}
\end{equation}\\

Thus, we see that without necessarily invoking the system perspective of maximizing the potential (which is equivalent to maximizing entropy with constraints), we can derive the Boltzmann distribution easily. This important property is not seen that clearly in statistical mechanics. In statistical mechanics, we typically maximize entropy or minimize Gibbs free energy to arrive at equilibrium results. That is, the emphasis is on the \textit{system perspective}; the individual agent's view is not given importance.\\

On the other hand, in statistical teleodynamics, the equivalence of the agent perspective and the system perspective is clearly seen. And the agent perspective becomes more and more valuable as we go up the capabilities spectrum, for the role of individuals and their "micro" properties and behavior become central to the analysis. For example, for rational economic agents (i.e., the 100\%-limit) that we consider next, defining the effective utility for an individual is paramount. Equally important is the recognition that the equilibrium is determined by the equality of utilities criterion, which is decidedly an agent-based view. \\

These valuable results, reproducing well-known equations in statistical thermodynamics, validate the soundness of statistical teleodynamics and its direct correspondence with statistical physics. A much more elaborate discussion on this connection and its implications can be found in \cite{venkat2017}.

\section*{Emergence of income distribution as a game theoretic outcome}

Having explored the 0\%-limit, let us now focus on the 100\%-"bookend", the limit of rational economic agents, somewhat like \textit{Homo Economicus}. Again, we summarize our earlier work here for the readers' convenience and to provide the overall context. At this limit, following  Venkatasubramanian~\cite{venkat2017}, we consider the emergent economic behavior of $N$ agents competing for jobs in an ideal free-market society. In our model, the effective utility, $h_{ij}$, enjoyed by an agent $i$ working in job $j$, earning a salary $S_j$ in a company is given by 

\begin{equation}
{
{h}_{ij}(S_j,N_j)=\alpha_i\ln S_j-\beta_i(\ln S_j)^2-\gamma_i\ln N_j\label{utility_job}}
\end{equation}\

where $\alpha_i,\beta_i,\gamma_i>0$ and  $N_j$ is the number {of} agents employed in job $j$, $j\in\{1,...,n\}$. We won't discuss how this model is arrived at here as it is not necessary for this paper's objectives. Furthermore, it is addressed at great length by  Venkatasubramanian~\cite{venkat2017}. Suffice to say that the first term on the right-hand side models the utility of salary, the second the disutility of effort expended to earn that paycheck, and the third the disutility of competition at that salary level. This effective utility is essentially the net value of the benefit from the paycheck after subtracting the costs of effort and competition. \\

In general, {$\alpha_i$, $\beta_i$ and $\gamma_i$,} which model the relative importance an agent assigns to the three terms in this equation, can vary from agent to agent. However, we discuss the ideal situation where all agents have the same preferences (i.e., a 1-class society), and hence treat these as constant parameters and drop the index $i$ to obtain the following equation:

\begin{equation}
{
{h}_{j}(S_j,N_j)=\alpha\ln S_j-\beta(\ln S_j)^2-\gamma\ln N_j\label{eq:utility_job_new}}
\end{equation} 

The question we are interested in is the following. Given a large collection of such agents competing for jobs in an ideal free-market society, where the agents are free to switch jobs in their pursuit of higher utilities and employers are free to hire/fire employees at will in their pursuit of higher profits, what will be the emergent collective behavior of such self-organizing dynamics?\\

Applying our game-theoretic analysis from above to this 100\%-limit, we can show (see \cite{venkat2015}, \cite{venkat2017}) that the potential $\phi(\mathbf{x})$ is given by

\begin{eqnarray*}
\phi(\mathbf{x})&=&\sum_{j=1}^n\int {h}_j(\mathbf{x}){d}x_j
\end{eqnarray*}

\begin{equation}
\phi(\mathbf{x}) = \phi_{u}+\phi_v+\phi_w+\text{constant}\label{pay_potential}
\end{equation}
where
\begin{eqnarray}
\phi_u&=&\alpha\sum_{j=1}^nx_j\ln S_j\\
\phi_v&=&-\beta \sum_{j=1}^nx_j(\ln S_j)^2\\
\phi_w&=&\frac{\gamma}{N} \ln \frac{N!}{\prod_{j=1}^n(Nx_j)!}\label{fair_potential}
\end{eqnarray}

We see that $\phi(\mathbf{x})$ is strictly concave:

\begin{eqnarray}
{\partial^2 \phi(\mathbf{x})}/{\partial x_j^2}=-{\gamma}/{x_j}<0
\end{eqnarray}

Therefore, as before, {\em a unique Nash Equilibrium} for this game exists, where $\phi(\mathbf{x})$ is maximized. Thus, the self-organizing free market dynamics ultimately reaches an equilibrium state with an emergent equilibrium income distribution. \\

Maximizing the potential in the Lagrangian framework as above, Venkatasubramanian et al. [2015] \cite{venkat2015} showed the emergent equilibrium is a lognormal distribution in income, given by

\begin{equation}
x_j^*=\frac{1}{S_jD}\exp\left[-\frac{\left(\ln S_j-\frac{\alpha+\gamma}{2\beta}\right)^2}{\gamma/\beta}\right]\label{logn_potential}
\end{equation}

where $D=N\exp\left[\lambda/\gamma-(\alpha+\gamma)^2/4\beta\gamma\right]$ and $\lambda$ is the Lagrange multiplier.\\

Alternatively, as we observed in the \textit{Thermodynamic Game}, we can derive the lognormal distribution by using the agent-based criterion for equilibrium as well. This perspective yields   

\begin{equation*}
{h}_{j} = h^* =\alpha\ln S_j-\beta(\ln S_j)^2-\gamma\ln N_j^*,  ~~j \in \{1,\dots,n\}
\end{equation*} 

where $h^*$ is the effective utility enjoyed by all the agents at equilibrium. By following a procedure similar to the one we used to derive Eq~\eqref{eq:boltzmann2} from \eqref{eq:equil-h}, we can derive the lognormal distribution given by Eq~\eqref{logn_potential}. We note again that the lognormal distribution is the isoutility curve for this game.\\

We wish to observe that even though our analysis is analogous to the \textit{Thermodynamic Game}, the resulting distribution is very different. For an ideal gas, the corresponding energy distribution is {\em exponential}, whereas the income distribution is {\em lognormal}. This is due to the differences in the objective functions and the constraints in the two systems. Generalizing this model for a 2-class society, i.e., with two different sets of $\alpha_i,\beta_i,\gamma_i$ values, the theory predicts the emergence of a lognormal plus power-law-like two-tier distribution that has been observed empirically. Since discussing such extensions would take us away from the objectives of this paper, we refer the reader to Venkatasubramanian~\cite{venkat2017}. \\

An important insight from our analysis, discussed at length in \cite{venkat2017} and \cite{venkatasubramanian2019much}, is the recognition that entropy really is a measure of \textit{fairness} in a distribution, not a measure of disorder as it has been understood conventionally for about 150 years. Thus, by maximizing entropy one is really maximizing the fairness in the allocation of income. When everyone is compensated as per the lognormal distribution, every worker enjoys the \textit{same effective utility} ($h^*$), thereby treating everyone equitably and hence fairly. This result answers a 200-year-old open question, since the days of Adam Smith, in political economy: \textit{How much income inequality is fair?} The answer is that the lognormal income distribution is the fairest distribution of income for a 1-class ideal free-market society~\cite{venkat2017}. Since an extensive discussion on the implications of this and other related results is not directly relevant to this paper's main objectives, we refer interested readers to Venkatasubramanian \cite{venkat2017} and \cite{venkatasubramanian2019much} for a comprehensive treatment. \\

\section*{Active brownian particles -- \textit{Chemotaxis game}}

Having demonstrated the validity and usefulness of our theory at the two limits,  we will now proceed to do the same for agents that are in between these limits.  We have deliberately chosen three entirely different kinds of active agents to demonstrate the wide applicability of our theory: (i) Active self-propelling particles in chemotaxis (\textit{Chemotaxis Game}) discussed here, (ii) Ants that ferry sand grains and create a crater (\textit{Ants Game}), and (iii) Segregation dynamics and phase equilibria among social agents (\textit{Schelling Game}). The last two are addressed in the following sections. \\ 

In chemotaxis, bacteria self-propel themselves towards (or away from) regions of higher concentration of some resource needed, a chemoattractant, or to be avoided for chemorepellant, for survival and growth. For the sake of simplicity and without any loss of generality, we only consider the chemoattractant case. The utility enjoyed by a bacterium increases with resource concentration ($c_i$), but is reduced by the disutility of competition ($-\ln N_i$) for that resource from other bacteria. This competition term models a bacterium's incessant search for a better utility to improve its survival fitness. This is an innate characteristic of all living systems as Darwin explained. \\

We model the utility of a state $i$ for a bacterium as:

\begin{eqnarray}
    h_i(c_i, N_i) = \alpha c_i - \ln N_i
\end{eqnarray}

where $c_i$ is the concentration of the resource in state $i$, and $N_i$ is the number of agents competing in that state. $\alpha$ denotes the affinity of the agent to the resource ($\alpha>0$ for chemoattractants and $\alpha<0$ chemorepellants). We only consider $\alpha>0$ as noted.\\

As before, the game-theoretic potential can now be determined using Eq~\eqref{eq:potential}, 
\begin{eqnarray*}
    \phi(\mathbf{x}) = \sum_{i=1}^n \int h_i(c_i, \mathbf{x})d x_i
\end{eqnarray*}

to arrive at 

\begin{eqnarray}
   \phi(\mathbf{x}) = \alpha \frac {C}{N} + \frac {1}{N} \ln \frac{N!}{\prod_{i=1}^n (Nx_i)!}
\end{eqnarray}

where $C$ is the total resource that is distributed over the entire state-space.\\

In the continuum limit, $c_i$ can be replaced by $c(\mathbf{r})$, the concentration distribution of the resource in the bacterial environment at location $r$. Similarly, $N_i$ is replaced by $\rho(\mathbf{r})$, the bacterial number density at a given location. Now the game theoretic potential becomes, 

\begin{eqnarray}
    \phi =  \frac{\int \alpha c(\mathbf{r}) \rho(\mathbf{r}) d\mathbf{r}}{N} - \frac{1}{N}\int \rho(\mathbf{r}) \ln \rho(\mathbf{r}) d \mathbf{r} \label{eq:potential-chemotaxis}
\end{eqnarray}

At any given time, an agent at location $\mathbf{r}$ will move to a new location $\mathbf{r}+\Delta \mathbf{r}$ if the utility at the new location, $h(\mathbf{r}+\Delta \mathbf{r})$, is greater than $h(\mathbf{r})$. The utility at the new location can be written in terms of the Taylor's series with respect to the old location as

\begin{eqnarray*}
    h(\mathbf{r}+\Delta \mathbf{r}) = h(\mathbf{r}) &+& (\Delta \mathbf{r}) \cdot \nabla h(\mathbf{r}) + \frac{1}{2}(\Delta \mathbf{r}) \cdot \nabla^2 h(\mathbf{r}) \cdot (\Delta \mathbf{r})\\
    &+& \text{ higher order terms } \dots
\end{eqnarray*}
At a sufficient level of granularity, since the agent cannot move a large distance in a single time step, we can omit the terms greater than order one. It is clear then that the agents will move in the direction of the gradient of utility to maximize their utility at any given time step. \\

The gradient of the utility acts like a self-imposed "force" on the agent -- i.e., its self-actualization --  in terms of the gradient of the number of particles in its immediate neighborhood -- i.e., the effect of local competition, and the gradient of the concentration of available resource -- i.e, the effect of increased resources. Therefore, the velocity of the agents $\mathbf{v}(\mathbf{r})$ (self-driven motion of the bacteria) at the location $\mathbf{r}$, in the absence of a convective field, is proportional to the gradient, which in turn is determined by the gradient in concentration of the resources and the number of agents around the location. This leads to

\begin{eqnarray}
    \mathbf{v}(\mathbf{r}) &\propto& \nabla h(\mathbf{r}) = \nabla \left(\alpha c(\mathbf{r})- \ln\rho(\mathbf{r})\right)\\
    \mathbf{v}(\mathbf{r}) &=& \zeta\nabla \left(\alpha c(\mathbf{r})- \ln\rho(\mathbf{r})\right)
\end{eqnarray}

where $\zeta$ is a constant parameter that reflects the ability of an agent to travel a discrete distance for a given value of gradient. As can be seen from the dimensions, $\zeta$ has a dimension of $length^2/time$. Low values of $\zeta$ make an agent travel short distances and high values long distances, in the same time interval. \\

{The continuity equation is a mass conservation equation, }
\begin{eqnarray}
    \frac{\partial \rho}{\partial t} = -\nabla\cdot \mathbf{J}(\mathbf{r}) \label{eq:continuity}
\end{eqnarray}

where the flux of the agents at a location, $\mathbf{J}(\mathbf{r}) = \rho(\mathbf{r}) \mathbf{v}(\mathbf{r})$,  is given by,

\begin{eqnarray}
    \mathbf{J}(\mathbf{r}) = \zeta \rho(\mathbf{r}) \nabla\left[\alpha c(\mathbf{r}) - \ln \rho(\mathbf{r})\right] \label{eq:active-flux}
\end{eqnarray}

We compare our analysis with that of O'Byrne and Tailleur \cite{o2020lamellar}, who proposed a coarse-grained diffusive model for active matter dynamics driven by a concentration field $c$. They describe the resultant coarse-grain dynamics using the following free energy functional $\mathcal{F}$ and deterministic flux $J_D$.
\begin{eqnarray}
\mathcal{F} &=  &\int d \mathbf{r} \rho(\mathbf{r}) \log \rho(\mathbf{r}) + \left[\frac{v_1}{v_0} + \frac{\alpha_1 + (d-1)\Gamma_1}{\alpha_0 + (d-1)\Gamma_0}\right]\frac{v_0}{d D} \nonumber \\
& &\times \frac{1}{p!}\int K(\mathbf{r}_1, \dots \mathbf{r}_p) \rho(\mathbf{r}_1) \dots \rho(\mathbf{r}_p) d \mathbf{r}_1 \dots \mathbf{r}_p \label{eq:fe-o2020}\\
J_D &= & -D \rho \nabla \left\{\left[v_1 + v_0\frac{\alpha_1 + (d-1)\Gamma_1}{\alpha_0 + (d-1)\Gamma_0}\right]\frac{c}{d D} + \log \rho \right\} \label{eq:jd-o2020}
\end{eqnarray}
where $J_D = \nabla \left(\delta\mathcal{F}/\delta \rho\right)$, the gradient of the functional derivative of the free energy ($\rho$ is the particle density).  The kernel term in the free energy (Eq~\eqref{eq:fe-o2020}) is the integral of the concentration field functional with the density across the phase-space.\\ 

We draw the reader's attention to the equivalence of these two approaches. Observe that our  Eq~\eqref{eq:potential-chemotaxis} and Eq~\eqref{eq:active-flux} map to Eq~\eqref{eq:fe-o2020}  and Eq~\eqref{eq:jd-o2020} with the recognition that  $\zeta = D$ and $\alpha= -\left[v_1 + v_0\dfrac{\alpha_1 + (d-1)\Gamma_1}{\alpha_0 + (d-1)\Gamma_0}\right]\dfrac{1}{dD}$. \\

Note also that minimizing free energy corresponds to  maximizing game theoretic potential. The simulations performed in their study (with the following parameter values: $v_0=1, v_1 = 0.2, \alpha_0 = 50, \alpha_1 = \Gamma_1=0$) correspond to $\alpha<0$ in our formulation, the chemorepellant case. \\

Several researchers have noted the surprising puzzle \cite{o2020lamellar, cates2013active, cates2015motility, gonnella2015motility} that the dynamics of these out-of-equilibrium active matter systems follow equivalent passive matter equilibrium dynamics based on free energy functional. Our framework resolves this puzzle by showing how active matter dynamics would lead to a game-theoretic \textit{arbitrage equilibrium} that is equivalent to a conventional statistical equilibrium. So, it is no surprise that such behavior is observed in many active matter systems. \\

We see that the free energy formulation is in fact the negative of the game-theoretic potential that the system tries to maximize. This in turn shows that chemical potential can be understood in terms of utility. For example, in \cite{cates2015motility}, the authors mention chemical potential as being $\mu = \ln v(\rho) + \ln \rho$. In our framework, this is equivalent to $h_i(v_i, N_i) = -\ln v_i -\ln N_i$. \\

\section*{Ant crater formation -- A statistical teleodynamics perspective}

\begin{figure}[!ht]
\centering
\includegraphics[width=0.5\linewidth]{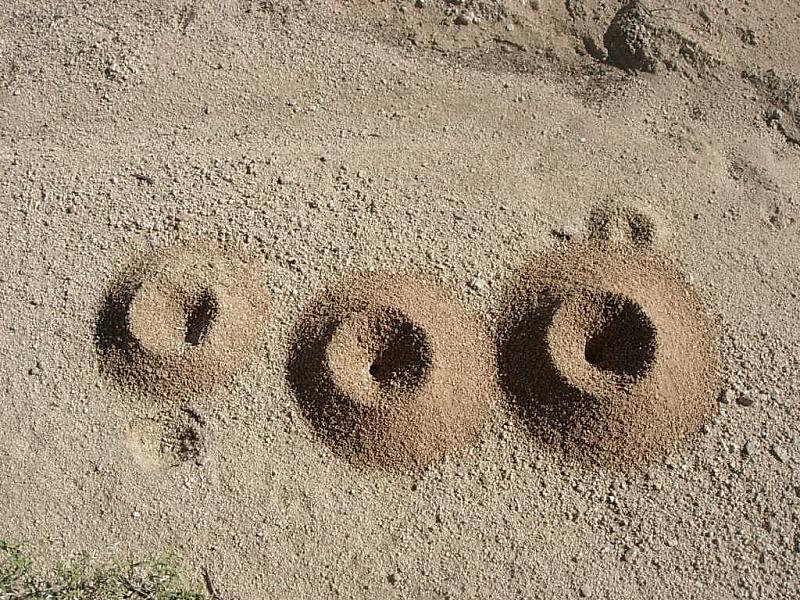}
\caption{{\bf The axisymmetric nature of ant craters}
(``Ant Hill Symmetry" by killkudzu is licensed under CC BY-NC 2.0, taken from {https://www.flickr.com/photos/22548183@N06/2249296478})}
\label{fig1}
\end{figure}

As another example of active matter in our agency spectrum, we now consider the collective behavior of ants in creating sand grain craters (Fig~\ref{fig1}). The dynamics of this activity involves transporting sand grains from an underground nest to above ground. Since transporting grains involves effort, which increases with the distance the ant travels, we suggest that the ants would prefer to drop the grains off sooner than later to minimize the effort. However, if they drop them off too close to the nest, then the sand grains pile could collapse back into the nest, which would mean more work for them later on. Thus, ants innately balance the need to transport the grains as far away as possible while trying to minimize the effort (i.e., the disutility) expended in doing so.\\ 

Therefore, in our model, an ant's utility is determined by three terms. The first term is the utility it gains by getting the grains out from below to form the nest, given by $b$. The second term describes the disutility it incurs by carrying the grains away from the nest. We assume that the ants move outward radially from the nest with some average velocity $v$. We model the rate at which the ants drop off the grains as $s r^{a-1}$, where $r$ is the distance it travels from the nest to the drop-off point ($s > 0$ and $a > 1$ are constant parameters). The disutility of effort $W$ an ant incurs in ferrying grains then depends on how much it carries and for how long. This results in,

\begin{eqnarray*}
    W = \int_0^t sr^{a-1} d t = \int_0^r sr^{a-1} \frac{d r}{v} = \frac{sr^a}{va} = \frac{\omega r^a}{a}
\end{eqnarray*}
where $\omega = s/v$. Those familiar with chemical engineering and reaction kinetics will notice that the quantity $W$ is, in fact, the Damk\"ohler number which quantifies the ratio of time scales of flow to the reaction. Higher values of Damk\"ohler number, in this case, suggest a higher propensity of an ant to drop off the grains. \\
 
The third term, $-\ln N_i$, accounts for the disutility of competition among the ants, similar to what we saw in the examples above. As the ants ($N_i$) try to crowd at the same location ($r_i$) to drop off their grains, this disutility term forces them to spread out to minimize the effect of competition.  \\ 

Combining all three, the total utility $h_i$ that an ant gains by dropping off a grain at a distance $r_i$ is given by,

\begin{equation}
    h_i(r_i, N_i) = b - \frac{\omega r_i^a}{a} - \ln N_i \label{eq:ant-utility}
\end{equation}

Note the structural similarity between the equations for the effective utility of ants (Eq~\eqref{eq:ant-utility}) and that of people hunting for jobs (Eq~\eqref{eq:utility_job_new}).\\

The potential for this system then becomes,

\begin{eqnarray}
    \phi(\mathbf{x}) &=& \sum_{i=1}^n \int h_i(\mathbf{x}) d x_i\\
    &=& b - \frac{\omega}{a} \langle r^a \rangle + \frac{1}{N} \ln \frac{N!}{\prod_{i=1}^N (Nx_i)!} 
\end{eqnarray}
where $\langle r^a\rangle$ is the expectation of the quantity $r^a$, based on the locations of the ants ($N$ is the total number of ants). \\

Similar to the \textit{Income Game}, it can be shown that $\partial^2 \phi/\partial x_i^2<0$, thereby proving a unique Nash equilibrium outcome for this emergent behavior. Furthermore, we can show using the Lagrange multiplier formulation, 

\begin{eqnarray}
x_i = \frac{\exp\left(-\dfrac{\omega r_i^a}{a} \right)}{\sum_j \exp\left(-\dfrac{\omega r_j^a}{a}\right)} \label{eq:ants}
\end{eqnarray}

In the continuum limit, the states are continuous, where  the state is defined as the radial location $r$. This results in the simplification $x_i = N_i/N =  \rho(r)2\pi r d r/N$, where $\rho(r)$ is the number density of the grain carrying ants at any location $r$. With this result in Eq~\eqref{eq:ants}, it can be shown that the number density of the ants follows the distribution, 

\begin{equation}
    \rho(r) = \frac{A}{r} \exp\left({-\frac{\omega r^a}{a} }\right)
    \label{eq:ants_2}
\end{equation}

where $A$ is a constant that satisfies the boundary condition of a constant flux of ants from the center of the nest. \\

Note that this emergent distribution is of the number of grain-carrying ants. Given this distribution of $\rho(r)$, the grain distribution can be calculated using a cumulative distribution

\begin{eqnarray*}
F(r) = \frac{\int_0^r (s r^{a-1} \rho) 2\pi r ~d r}{\int_0^{{\infty}} (s r^{a-1} \rho) 2\pi r ~d r}
\end{eqnarray*}

This gives,
\begin{eqnarray*}
    F(r) = 1 - \exp{\left(- \frac{\omega r^{a}}{a} \right)}
\end{eqnarray*}
with the grain distribution $f(r)$, given by:
\begin{eqnarray}
    f(r) = \frac{d F}{d r} = \omega r^{a-1} \exp\left(-\frac{\omega r^{a}}{a}\right)
\end{eqnarray}

which turns out to be the Weibull distribution. \\

We compare our game-theoretic analysis with that of Picardo and Pushpavanam (2015) \cite{picardo2015understanding}, who invoked a fluid mechanics analogy for the flow of ants. They formulated a continuum model using the steady state governing equation,
\begin{eqnarray*}
    v \ \rho(r)\ 2\pi r - v \ \rho(r+\Delta r)\  2\pi (r+\Delta r) &\nonumber\\
    -
    \int_r^{r+\Delta r} (2\pi r') \ K \rho(r') ~ d r' &= 0
\end{eqnarray*}
where $v$ is the rate at which ants leave the nest, $\rho(r)$ is the number density of ants at $r$, and $K$ is the rate at which each ant drops a grain of sand. This, in the limit of $\Delta r\rightarrow 0$, gives,
\begin{eqnarray*}
    v \left(\frac{1}{r} \frac{d}{d r} (r \rho)\right) = -K\rho
\end{eqnarray*}
Assuming a power law dependence of $K$ on $r$, i.e., $K = sr^{a-1}$. we have 
\begin{eqnarray*}
    v \left(\frac{1}{r} \frac{d}{d r} (r \rho)\right) = -sr^{a-1} \rho
\end{eqnarray*}

Defining $\omega = s/v$, we get the general solution,

\begin{equation}
    \rho(r) = \frac{A}{r} \exp\left({-\frac{\omega r^a}{a} }\right)
\end{equation}

which is the same as our Eq~\eqref{eq:ants_2} and with the same resultant Weibull distribution for the grains. \\

\section*{Phase equilibria in social segregation dynamics -- Schelling game}
As the last example, we consider phase transitions in active matter systems, the segregation dynamics of a Schelling-like model. In this model, two classes of agents with different utility preferences, which depend on the interactions with their neighbors, switch neighborhoods to increase their utilities, whenever possible. Schelling showed that having agents with even "weak" in-class preference towards their own kind could still lead to a highly segregated society \cite{schelling1971dynamic}. \\

Many researchers have analyzed this model from the perspective of statistical mechanics {and population dynamics} \cite{odor2008self, muller2008inhomogeneous, stauffer2008social, gauvin2009phase, grauwin2009competition, sienkiewicz2009nonequilibrium, chuang2019network}. We compare our analysis with that of Grauwin et al.~\cite{grauwin2009competition}. They first present the formulation for a system of agents and empty sites on a lattice, where each agent prefers not to be lonely but among many others. However, they also would like to avoid overcrowding. \\

Like the other examples above, we formulate this problem by first defining the effective utility, $h_i$, for the agents, which the agents try to maximize by switching to better neighborhoods. As before, our effective utility is the net sum of benefit minus cost. The benefit component has two terms. First, since the agents prefer to have more neighbors, this \textit{affinity benefit} term is proportional to the number of agents in its immediate neighborhood. Second, the agents also derive a benefit by having a large number of vacant sites to potentially switch to in the future should such a need arise (i.e., the \textit{option benefit} term). Similarly, we have two terms on the cost or disutility side also. One is the day-to-day disutility of the hassle of dealing with lots of neighbors - this is the disutility of overcrowding (i.e., the \textit{congestion cost} term). The second term is the disutility due to competition for desirable future sites to switch to (i.e., the \textit{competition cost} term). \\

Following Grauwin et al.\cite{grauwin2009competition}, we consider a lattice in terms of local neighborhoods or blocks, each with $H$ sites which the agents can occupy. There are $n$ blocks and $N$ agents, with an average agent density of $\rho_0 = N/(n H)$. The state of the system is defined by specifying the number of agents, $N_i$, in block $i$ for all blocks ($i \in \{1, \dots, n\}$ ). Block $i$ has $V_i$ vacant sites, so  $V_i = H - N_i$.  We model the effective utility of state $i$ as, 

\begin{eqnarray}
    h_i(N_i) = \eta N_i+ \ln(H - N_i) - \xi N_i^2 - \ln(N_i) 
\end{eqnarray}
 
The first two terms are the benefits and the last two are the costs of an agent's interactions with its neighbors and vacant sites, where $\eta, \xi >0$. Rewriting this in terms of the fraction of agents in state $i$, $x_i=N_i/N$, we have

\begin{eqnarray*}
    h_i(\mathbf{x}) = \eta N x_i -\xi N^2 x_i^2 - \ln x_i + \ln \left(\frac{H}{N} - x_i\right)
\end{eqnarray*}

or in terms of the density of agents in state $i$, $\rho_i = N_i/H$, 

\begin{eqnarray}
    h_i(\boldsymbol{\rho}) = \eta H \rho_i - \xi H^2 \rho_i^2 - \ln\rho_i  + \ln (1-\rho_i)
\end{eqnarray}

For the sake of simplicity, we rewrite $u(\rho_i) = \eta H \rho_i - \xi H^2 \rho_i^2$, with $\rho_i = x_i(N/H)$). It can be seen that 

\begin{eqnarray}
    \phi(\mathbf{x}) &=& \sum_{i=1}^n \int h_i(\mathbf{x}) d x_i = \frac{H}{N}\sum_{i=1}^n \int h_i(\boldsymbol{\rho}) d \rho_i \nonumber\\
    &=&\frac{H}{N}\sum_{i=1}^n \int_0^{\rho_i}\left[ u(\rho) - \ln \rho + \ln (1-\rho)\right] d \rho \nonumber\\
    &=&\frac{H}{N}\sum_{i=1}^n  - \rho_i \ln \rho_i - (1-\rho_i)\ln (1-\rho_i) + \int_0^{\rho_i}  u(\rho)d\rho 
    \label{eq:schelling-potential}
\end{eqnarray}

As before, to analyze the equilibrium behavior, we don't necessarily need to maximize $\phi (x)$  or minimize the free energy as Grauwin et al.\cite{grauwin2009competition} do. We can take the simpler agent-based perspective and exploit the fact that at equilibrium all agents have the same utility, i.e.,

\begin{eqnarray}
    \eta N_i^* - \xi (N_i^*)^2 - \ln (N_i^*) + \ln (H-N_i^*) = h^* \nonumber\\
    (\eta H) \rho^* - (\xi H^2) (\rho^*)^2 - \ln(\rho^*) + \ln (1-\rho^*) = h^* \label{eq:schelling-eqbrm}
\end{eqnarray}

\begin{figure}[!ht]
    \centering
    \includegraphics[width=\linewidth]{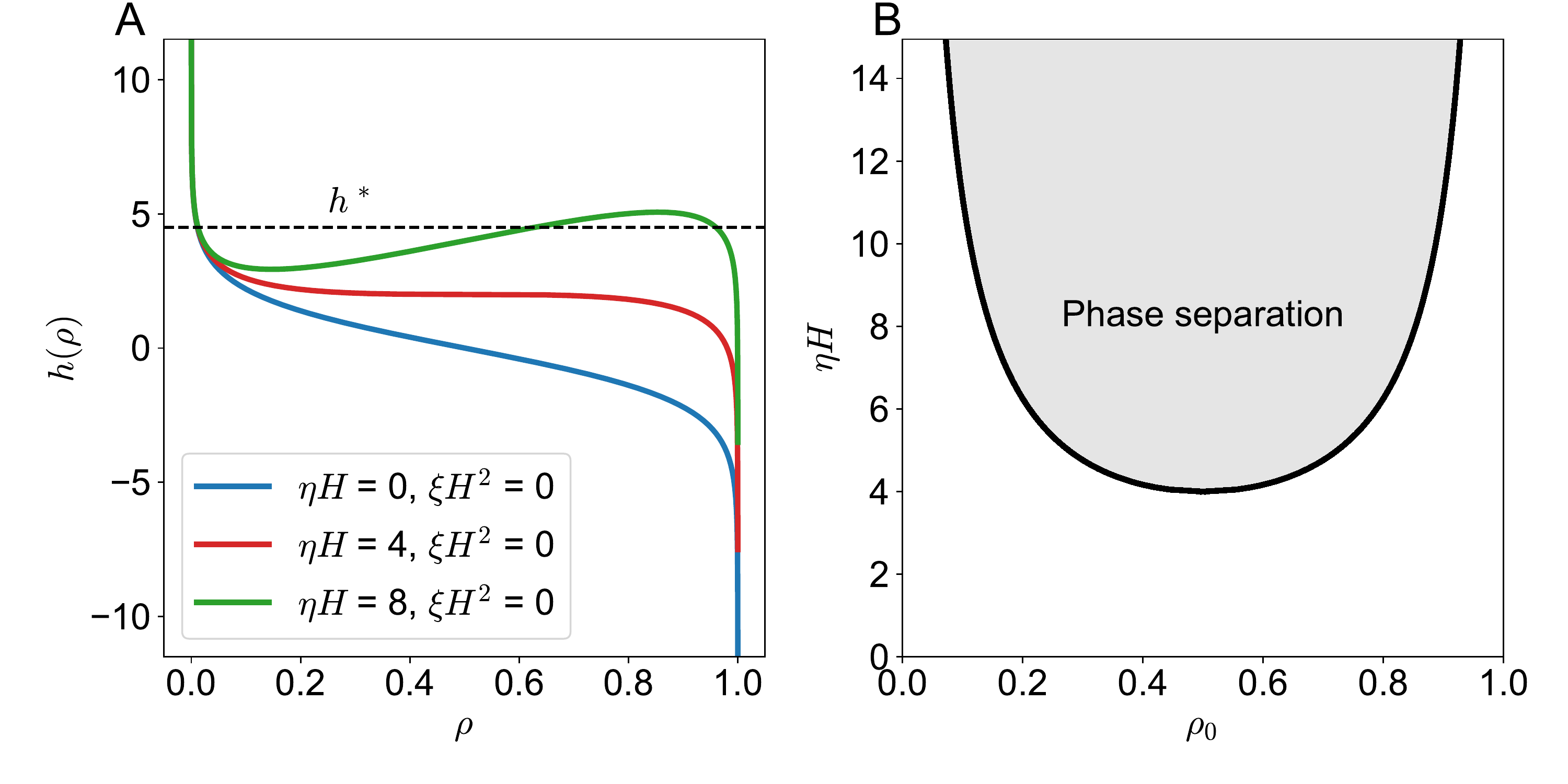}
    \caption{{\bf  Phase separation is governed by the benefit-cost trade-offs}
    $h^*$,  Eq~\eqref{eq:schelling-eqbrm}, depends on the parameters $\eta H$ and $\xi H^2$. A: Effective utility $h^*$ as a function of the block density $\rho$ for $\xi=0$. B: Gray region is where  $\partial h/\partial \rho\big\vert_{\rho^*}>0$ for $\xi=0$. Phase separation occurs for values of the average density $\rho_0$ and $\eta H$ within this region.}
    \label{fig2}
\end{figure}

We explore the behavior of $h^*$ as a function of $\rho ^*$ (Eq~\eqref{eq:schelling-eqbrm}) numerically as shown in Fig~\ref{fig2}. For the sake of simplicity we set $\xi = 0$. This can be relaxed readily and the outcome would be the same qualitatively. We observe that for a given $h^*$, we have a unique solution ($\rho ^*$) when $\eta H = 0$ (blue curve). However, for higher values of $\eta H$, we could obtain multiple densities for the same $h^*$. These can be divided into low density, intermediate density, and high-density blocks. The intermediate density solution, however, is unstable as $\partial h/\partial \rho\rvert_{\rho^*}>0$. \\

Therefore, for high values of $\eta H$ (e.g., green curve), we observe the spontaneous emergence of two phases, high and low density of agents, at equilibrium. Intuitively, in the high-density blocks, the agents derive so much more benefit from the affinity term (due to high $\eta H$) that it more than offsets the disutilities due to congestion and competition, thereby yielding a high effective utility. Similarly, in the low-density blocks, the benefits of low congestion and competition combined with increased options more than compensates for the loss of utility from the affinity term. Thus, every agent enjoys the same effective utility $h^*$ in one phase or the other at equilibrium. This causes equilibrium because, as noted, there is no more arbitrage incentive left for the agents to switch states. \\

As noted, the stability of the phases is determined by $\partial h/\partial \rho\big\vert_{\rho^*}$ given by ($\xi = 0$ in Eq~\eqref{eq:schelling-eqbrm})

\begin{eqnarray}
    \frac{\partial h}{\partial \rho}\bigg\vert_{\rho^*} = \eta H - \frac{1}{\rho^*} - \frac{1}{1-\rho^*}
\end{eqnarray}

Onset of multiple solutions is established when this derivative is zero, that is, at  $\rho ^*(1-\rho ^*) = 1/(\eta H)$. This yields  $\rho_{\pm} = (1 \pm \sqrt{1-4/(\eta H)})/2$ and two phases will form when $\eta H>4$. This is also seen in Fig~\ref{fig2}{A}, where at $\eta H = 4$ (red curve) the curve flattens at $\rho=0.5$. At $\eta H = 8$, $\rho_+ = 0.854, ~\rho_- = 0.146$. In Fig~\ref{fig2}{B}, we show the equilibrium phase separation region. Readers will no doubt recognize the striking similarity of this figure with a  spinodal decomposition phase diagram~\cite{cahn1961spinodal, favvas2008spinodal}. This congruence should not be surprising as the underlying arbitrage dynamics mechanism for both phenomena is essentially the same as our theory demonstrates. \\

\begin{figure}[!ht]
    \centering
    \includegraphics[width=\linewidth]{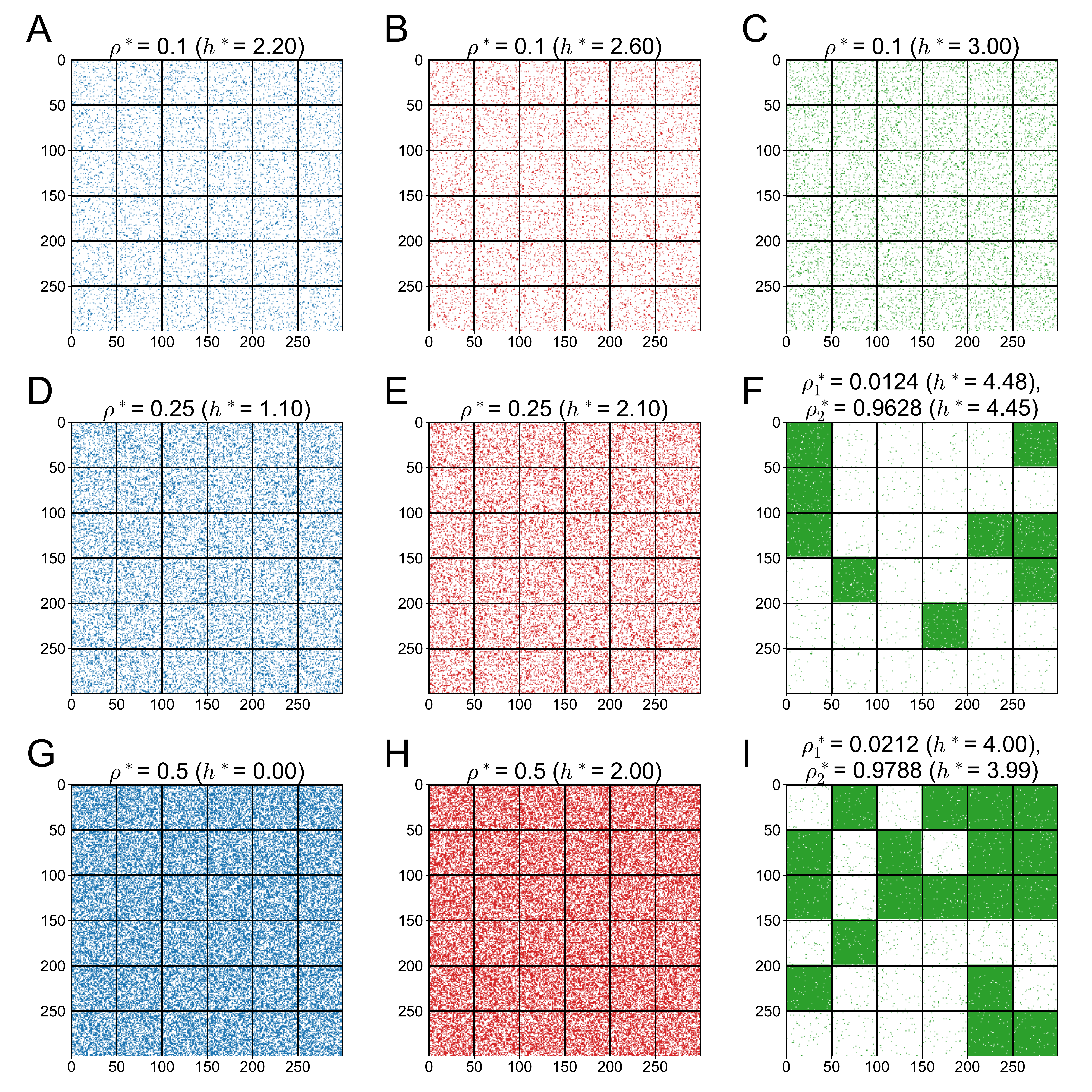}
    \caption{ {\bf Equilibrium configuration after one million time steps on a $300\times 300$ grid, $H=50\times 50$ and $\xi=0$, for different $\rho_0$ and $\eta H$}. 
    Top row(\textit{A, B, C}), middle row(\textit{D, E, F}), and bottom row(\textit{G, H, I}) correspond to an average density of $\rho_0 = 0.1, 0.25, 0.5$, respectively. Left column (\textit{A, D, G}), middle column (\textit{B ,E, H}), and right column (\textit{C ,F, I}) correspond to $\eta H=0, 4, 8$, respectively. For single-phase systems, the final density is seen to be the same as $\rho_0$. For multiphase systems the density of the two phases are \textit{F.} 0.0124 and 0.9628 (utility of 4.48, 4.45, respectively) \textit{I.} 0.0212 and 0.9788 (utility of 4.00, 3.99 respectively)}
    \label{fig3}
\end{figure}

We experimentally validated the phase segregation using an agent-based simulation on a $300\times 300$ grid (90,000 sites total), with block sizes of $H=50\times 50$ sites (2500 sites), for different $\rho_0$ ($\rho_0$ = 0.1 corresponds to $N$ = 9,000 agents; $\rho_0$ = 0.25, $N$ = 22,500; $\rho_0$ = 0.5, $N$ = 45,000) and $\eta H$ ($\xi=0$), as shown in (Fig~\ref{fig3}). For all configurations except (F) and (I) in Fig~\ref{fig3}, we see the emergence of single phase equilibria. For two-phase equilibrium, the density of the two phases are $\rho_1^*$ = 0.0124 and $\rho_2^*$ = 0.9628 with the corresponding effective utilities ($h^*$) of 4.48 and  4.45, respectively (Fig~\ref{fig3}{F}). In Fig~\ref{fig3}{I}, the two phases are $\rho_1^*$ = 0.0212 ($h^*$ = 4.00) and $\rho_2^*$ = 0.9788 ($h^*$ = 3.99). In (F) and (I), the negligible differences in $h^*$ values are due to the small inherent fluctuations in the equilibrium configurations. \\

Note the similarity of this game with the \textit{Chemotaxis Game} and also with the motility-induced phase separation (MIPS) dynamics discussed by Cates and Tailleur \cite{cates2015motility}. In MIPS, as we observed in the section on the \textit{Chemotaxis Game}, the effective utility is given by $h(\rho_i) = -\ln v(\rho_i) - \ln \rho_i$. Recall that the particle velocity $v(\rho_i)$ decreases with increased density $\rho_i$. In MIPS, for high values of diffusivity and velocities, two phases - high and low density - are observed. As we explained above, intuitively, in the high-density phase, the increased disutility of competition (i.e., $\ln \rho_i$ is high, here $\rho_i$ is not bounded by $<1$) is offset by the decreased disutility due to low velocity (in $\ln v(\rho_i)$). The opposite happens in the low-density phase. \\

For the block Schelling model, Grauwin et al. \cite{grauwin2009competition} formulate the potential function $F(\boldsymbol{\rho})$ for the configuration $\boldsymbol{\rho}$ as

\begin{eqnarray}
F(\boldsymbol{\rho}) = H\sum_{q}f(\rho_q)
 \label{eq:grauwin-F}
\end{eqnarray}

where $q$ is the block identity. The stationary distribution follows the probability $\Pi(x) = \exp(F(x)/T)/Z$. The block potential $f(\rho)$ is given by

\begin{eqnarray}
f(\rho) = -T\rho \ln \rho - T(1-\rho)\ln (1-\rho) + \alpha \rho u(\rho) \nonumber&\\ + (1-\alpha)\int_0^\rho u(\rho') d \rho' \label{eq:grauwin-f}
\end{eqnarray}

where $\alpha$ is a trade-off parameter of an agent's strategy to move to a new block. For  $\alpha=0$, the agent selfishly moves to a new block if it increases its utility. For  $\alpha=1$, the agent moves to the new block if the collective utility of all agents increases even if its own utility decreases. In our formulation, for the sake of simplicity, we only consider the $\alpha=0$ case. Again, this restriction can be relaxed without any change in the qualitative outcomes - it only makes the analysis more complicated and the comparison less transparent.\\

Observe that our game-theoretic potential $\phi (x)$ (Eq~\eqref{eq:schelling-potential}) is equivalent to their potential function $F(x)$ in  Eq~\eqref{eq:grauwin-F} and Eq~\eqref{eq:grauwin-f} with $\alpha=0$ (except for the scaling factor of $1/NT$). The effect of temperature $T$ in their formulation is captured in our parameters $\eta$ and  $\xi$. For instance, increasing $T$ in Eq~\eqref{eq:grauwin-f} is equivalent to decreasing $\eta$ or increasing $\xi$ in Eq~\eqref{eq:schelling-eqbrm}. That is, phase separations are unlikely when the benefit of affinity is low in comparison with the higher disutility of congestion. The profile of $u(\rho)$ is also similar in both models, an increase followed by a decrease as the trade-off between the affinity and congestion terms. Their $u(\rho)$ follows an inverted-V profile whereas ours is an inverted-U.\\

Grauwin et al. \cite{grauwin2009competition} also discuss a similar formulation for a Schelling game with two different classes of agents, $R$ and $G$, with individual densities $\rho_R, \rho_G$, respectively. The functional $f(\rho)$ in Eq~\eqref{eq:grauwin-f} generalizes to a new functional $f(\rho_R, \rho_G)$ given by

\begin{eqnarray}
    f(\rho_R, \rho_G) = &-&T\rho_R\ln \rho_R - T\rho_G\ln \rho_G \nonumber \\ &-& T (1-\rho_R -  \rho_G)\ln(1-\rho_R - \rho_G) \nonumber\\ 
    &+& \alpha \left[\rho_R u_R(\rho_R) + \rho_G u_G(\rho_G)\right] \nonumber\\
    &+&(1-\alpha)\left[\int_0^{\rho_R} u_R(\rho') d \rho'+\int_0^{\rho_G} u_G(\rho') d \rho'\right]
\end{eqnarray}

Likewise, in our framework, Eq~\eqref{eq:schelling-eqbrm} readily generalizes to (for $\alpha=0$)

\begin{eqnarray*}
    h_{R,i}^* &=& u_{R}(\rho_{R, i}^*) -\ln (\rho_{R, i}^*) + \ln (1 - \rho_{R, i}^* - \rho_{G, i}^*)\\
    h_{G,i}^* &=& u_{G}(\rho_{G, i}^*) -\ln (\rho_{G, i}^*) + \ln (1 - \rho_{R, i}^* - \rho_{G, i}^*)
\end{eqnarray*}
with the appropriate segregation dynamics and phase equilibria results. 

{Here, we would like to note the striking similarity between the Schelling game potential Eq~\eqref{eq:grauwin-f} and the Flory-Huggins theory \cite{flory1942thermodynamics, huggins1941solutions} for liquid-liquid phase segregation, where the volume fraction of the polymer in a solvent ($\rho$), affects the volumetric free energy ($f$) in a similar way. 
\begin{eqnarray}
\frac{f}{k_B T} = \chi \rho (1-\rho) + \frac{\rho}{N}\ln \rho + (1-\rho)\ln(1-\rho) \label{eq:fh}
\end{eqnarray}}

{where $N$ is the total number of monomers in the polymer molecule, and $\chi$ is the Flory-Huggins interaction parameter which accounts for solvent polymer interactions. Due to presence of multiple monomer units, we see an asymmetry in the entropy component, in contrast to the Schelling model formulation. Note how if we have one monomer unit, Eq~\eqref{eq:fh}, is comparable to Eq~\eqref{eq:grauwin-f}.}

{The free-energy, as stated earlier, is the negative of the chemical potential. Given the free-energy, we can derive the utility of a given volume fraction as, 
\begin{eqnarray*}
h(\rho) = \frac{\partial \phi}{\partial\rho} = -\frac{\partial f}{\partial \rho} = -(\chi+\frac{1}{N}+1) + 2\chi\rho - \frac{1}{N}\ln \rho + \ln (1-\rho)
\end{eqnarray*}}

{The initial factor however, is linear in $\rho$, saying that the utility in having a volume fraction $\rho$ increases with increasing volume fraction if the Flory-Huggins interaction parameter is positive.}

\section*{Discussion and conclusion}

Recent interest in active matter has brought to center stage a fundamental question that hasn't garnered much attention in mainstream physics for decades, except for an occasional paper [e.g., \cite{anderson1972more}]: What are the fundamental principles and the relevant mathematical framework for predicting and explaining the "macroscopic" properties and emergent behavior of a system comprising millions of goal-driven agents given their "microscopic" properties? In other words, how do we go from the parts to the whole for active matter? \\

Going from the parts to the whole is the \emph{opposite} of the \emph{reductionist} paradigm that has dominated the 20\textsuperscript{th} century scientific thinking. In the reductionist paradigm, one tries to predict and explain macroscopic properties by seeking deeper, more fundamental principles and mechanisms that cause such properties and behaviors. Reductionism is a \emph{top-down} framework, starting at the macro-level and then going deeper and deeper, seeking explanations at the micro-level, nano-level, and even deeper levels - i.e., \emph{from the whole to the parts}. Physics and chemistry were spectacularly successful in pursuing this paradigm in the last century, giving birth to quantum mechanics, quantum field theory, string theory, and so on. Even biology pursued this paradigm to phenomenal success and showed how heredity could be explained by molecular structures such as the double helix.  Some 600+ Nobel Prizes have been awarded over the last 120 years for pursuing the reductionist paradigm.\\ 

However, many grand challenges we face in 21\textsuperscript{st} century science are \emph{bottom-up} phenomena, going from the parts to the whole. Examples include predicting phenotype from genotype, predicting the effects of human behavior on global climate, and predicting the emergence of consciousness and self-awareness, \emph{quantitatively} and \emph{analytically}. By its very nature, reductionism cannot help here because addressing this challenge requires the \textit{opposite} approach - i.e., putting things together, \textit{not taking things apart}. For this, we need a novel systems engineering-like perspective.

Besides, reductionism typically doesn't concern itself with \emph{teleology} or \emph{purposeful} behavior. In contrast, modeling bottom-up phenomena requires addressing this vital feature as teleology-like properties often emerge at the macroscopic levels, either explicitly or implicitly, even in the case of purpose-free entities. So, we need a new paradigm, a bottom-up analytical framework - a \emph{constructionist} or \emph{emergentist} approach. \\

Physicists have been chasing for nearly a century the elusive grand unified theory of the four fundamental forces. Such a unified theory again is from the top-down reductionist perspective. \textit{But how about a grand unified theory from the bottom-up constructionist perspective,  a quantitative theory of emergent phenomena?} In this paper, we propose statistical teleodynamics as a step in that direction. By a bottom-up emergentist  framework, we do not mean the mere discovery of hidden patterns, such as in the performance of a "deep learning" neural network that discerns complex statistical correlations in vast amounts of data. We mean the need for a comprehensive mathematical framework that can explain and predict macroscopic behavior and phenomena from a set of fundamental principles and mechanisms. A theory that can not only predict the important qualitative, and quantitative, emergent macroscopic features, but can also \textit{explain} why and how these features emerge and why not some other outcomes. The current "deep learning" AI systems lack this capability. \\  

Current attempts towards a theory of active matter utilize concepts and methods from hydrodynamics, kinetic theory, and non-equilibrium statistical physics, but they don't address the central feature of \textit{purposefulness} in teleological systems.  We believe that this critical feature of purpose or goal, which in our opinion is the crucial difference between active and passive matter, has to be explicitly accounted for in any comprehensive theory of active matter. Of course, it is true that the survival purpose is absent in certain lower-forms of active matter.  However, they behave in a manner that seems like goal-seeking and so could be treated as such as we showed in our results above.\\

Statistical teleodynamics acknowledges the importance of recognizing the individual active agent and its "microscopic" behavioral properties explicitly in developing a bottom-up analytical framework of emergent phenomena. It also accounts for the role of purpose, and the naturally attendant concept of the \textit{pursuit of maximum utility}, overtly. This is accomplished by a unifying synthesis of potential game theory and statistical mechanics. This grand synthesis can be seen as the natural generalization of statistical thermodynamics for active matter or as the natural extension of potential game theory for passive agents. The theory considers a spectrum of self-actualizing capabilities, going from none to completely rational, with the various types of active matter occupying someplace in this spectrum.\\ 

Since our theory is a bottom-up emergentist framework, it \textit{emphasizes the agent perspective} in contrast with statistical mechanics that stresses the \textit{system} view. For example, whenever one discusses equilibrium in statistical mechanics, one usually formulates it in terms of maximizing entropy or minimizing free energy, which is a \textit{system} perspective. However, in statistical teleodynamics, while the system view is also present via the maximization of game-theoretic potential, the \textit{equality of effective utility} for all agents as the equilibrium criterion from the \textit{agent perspective} is conspicuously recognized and exploited. Accordingly, we explain that the arbitrage equilibrium is reached because the agents no longer have an incentive to switch states. The \textit{fundamental quantity} in our theory is the effective utility of an agent. Observe that for all the five cases presented, the first step is to formulate the equation that models the agent in terms of its effective utility ($h_i$), which of course is an agent property, not a system one. We formulate $h_i$ first and then derive $\phi (x)$ - i.e., we go from the parts to the whole, whereas in statistical mechanics one formulates Gibbs free energy first. \\ 

At the risk of belaboring this point, we must nevertheless emphasize the importance of two key philosophical differences between these two frameworks. First, while both statistical mechanics and statistical teleodynamics connect the parts and the whole - the microscopic and the macroscopic - the former takes the top-down view and starts with system-level properties such as Gibbs free energy to proceed. The latter, on the other hand, takes the bottom-up view and starts with the part-level property, effective utility, and then builds up. Second, the explicit recognition of purpose and utility. Both  philosophical differences become more crucial as we go up the hierarchy of self-actualizing capabilities with the agents becoming more and more autonomous and strategic.\\

Many researchers have noted the puzzling surprise \cite{berkowitz2020active,o2020lamellar, cates2013active, cates2015motility, gonnella2015motility} that some active matter systems that look like dynamic, out-of-equilibrium systems at the microscopic scale, behave macroscopically like simple equilibrium systems of passive matter. Our theory resolves this puzzle and explains why and how this behavior emerges naturally for both passive and active matter using the same analytical framework. Thus, all these different kinds of agents, possessing varying degrees of agency from none to perfectly rational, belong to the same \textit{universality class}. As our theory demonstrates, the main requirement to belong to this universality class is that, in the expression for an agent's effective utility, the disutility due to competition can be modeled (or at least, reasonably approximated) as $-\ln N_i$ (discrete case) or  $-\ln \rho$ (continuous case). This is a critical requirement as Kanbur and Venkatasubramanian explain~\cite{kanbur2020occupational}. This agent-based property directly leads to the system-wide property of entropy, thereby connecting the agents and the system in a cohesive mathematical framework. This term also facilitates the integration of potential game theory with statistical mechanics, thus paving the way for a universal theory of emergent equilibrium phenomena in active and passive matter. \\

Commenting on this point some more, observe the universality of the structure of the effective utility models for all the five very different kinds of agents. The \textit{disutility due to competition} is present in all of them. Furthermore, the other terms model the benefits and costs in terms of variables and parameters that are appropriate for the application domain. For instance, in the \textit{Thermodynamic Game}, the relevant quantities are energy and temperature, whereas in chemotaxis they are concentration gradient, particle velocity, and particle density, and in the \textit{Income Game} it is salary. Thus, the universal structure accommodates domain-specific concepts, mechanisms, and quantities. There is no need to artificially invoke concepts such as temperature and free energy, which make perfect sense in physical systems but are meaningless and awkward  in social and economic systems. This awkwardness arises when one tries to graft the system-view of statistical mechanics, with the attendant concepts such as Gibbs free energy and temperature, on to a system where these variables and parameters have no meaning \textit{per se}. This is avoided or greatly minimized when one takes the agent view, as one is now forced to reckon with concepts and variables/parameters that are appropriate and natural for agent-level interactions. For example, in Eq~\eqref{eq:utility_job_new} for the income game or Eq~\eqref{eq:schelling-eqbrm} for the Schelling game, there is no place for temperature. The agent-level dynamics and mechanisms of economic or social agents simply wouldn't allow such irrelevant quantities to be invoked. Thus the agent-based view enforces certain degree of conceptual clarity in the problem formulation. \\

As a result, our theory uses concepts and variables that are natural for the problem, making the modeling more transparent and intuitive. This makes it easier to understand and explain the emergent macroscopic behavior in terms of its microscopic properties. This facility is often missing in the top-down systems-view that one sees in most purely statistical mechanical formulations that use free energy minimization.  \\

Just as mechanical equilibrium is reached when the forces balance each other equally, and phase equilibrium is achieved when the chemical potentials are equal, our theory demonstrates that a system of active agents will reach \textit{game-theoretic} or \textit{arbitrage equilibrium} when their \textit{effective utilities are equal}.  Whenever the effective utility is of the form we saw in the five examples (mainly, the $-\ln N_i$ or $-\ln \rho$ term), this game-theoretic Nash equilibrium is equivalent to the statistical Boltzmann equilibrium. In fact, our theory reveals the critical insight that both active and passive agents are driven by arbitrage opportunities towards equilibrium, except that their \textit{arbitrage currencies} are different. For passive matter, the currency is chemical potential, whereas for active matter, effective utility. \\


In summary, statistical teleodynamics is a mathematical framework that unifies two seemingly disparate domains - purpose-free entities in statistical mechanics and purposeful agents in game theory - within the same conceptual formalism to predict and explain certain types of emergent equilibrium phenomena in active and passive matter systems. Besides the physics of active matter, this theory has interesting implications in political economy, sociology, and the organization of complex teleological networks.
We find this universality of bridging across seemingly disparate disciplines quite appealing and encouraging as we pursue the emergentist paradigm to connect the parts to the whole. This could be seen as a starting point for a more comprehensive analytical theory of design, control, and optimization through self-organization. 



\section*{Acknowledgments}
The authors would like to thank Jessica Shi for her contribution to the agent-based simulation of the segregation dynamics model. This work is supported in part by the Center for the Management of Systemic Risk (CMSR) at Columbia University.

\section*{Author contributions}
{VV developed the conceptual framework and its mathematical formulation. VV, AS, and LD performed the analysis. AS and LD performed the computer simulations. VV wrote the paper with input from AS and LD.}
\nolinenumbers

\bibliographystyle{model1-num-names}

\end{document}